\documentclass[12pt]{elsarticle}
\usepackage{amsmath}
\usepackage{amssymb}
\usepackage{graphicx}

\DeclareMathSymbol{\lang}{\mathord}{symbols}{"68}
\DeclareMathSymbol{\rang}{\mathord}{symbols}{"69}
\DeclareMathSymbol{\openbra}{\mathord}{symbols}{"68}
\DeclareMathSymbol{\closeket}{\mathord}{symbols}{"69}
\newcommand{\ket}[1]{{| #1 \closeket}}

\renewcommand{\phi}{\varphi}
\renewcommand{\epsilon}{\varepsilon}
\DeclareMathOperator{\Rre}{Re}
\DeclareMathOperator{\Iim}{Im}
\begin{document}
\begin{frontmatter}
\title{Matter-wave chaos with a cold atom in a standing-wave laser field}
\author{S.V.~Prants}
\ead{prants@poi.dvo.ru, tel.007-4232-312602, fax 007-4232-312573}
\address{Laboratory of Nonlinear Dynamical Systems,
Pacific Oceanological Institute of the Russian Academy of Sciences, 
Baltiiskaya St., 43, 690041 Vladivostok, Russia}
\begin{abstract}
Coherent motion of cold atoms in a standing-wave field is interpreted as a  
propagation in two optical potentials. 
It is shown that the wave-packet dynamics can be either regular 
or chaotic with transitions between these potentials after passing 
field nodes. Manifestations of de Broglie-wave chaos are found in the behavior of the 
momentum and position probabilities  
and the Wigner function. The probability of those transitions depends on the 
ratio of the squared detuning to the Doppler shift and is large
in that range of the parameters where the classical motion is shown to be chaotic.
\end{abstract}
\end{frontmatter}

\section{Introduction}

Cold atoms in a standing-wave laser field \cite{Minogin,K90,Meystre,Schleich,
Fedorov} are ideal objects for studying
fundamental principles of quantum physics, quantum-classical correspondence,
and quantum chaos. The proposal \cite{GSZ92} to study atomic dynamics in
a far-detuned modulated standing wave made atomic optics 
a testing ground for quantum chaos. A number of impressive experiments have
been carried out in accordance with this proposal \cite{Raizen,HH01,SW05}.
Furthermore, cold atoms have been recently used to
study different phenomena in statistical physics  
including ratchet effect of a directed transport of atoms 
in the absence of a net bias \cite{ML99,SS03,CB06} 
and Brillouin-like propagation modes in optical lattices \cite{SC02}.

New possibilities are opened if one works near the atom-field resonance where
the interaction between the internal and external atomic degrees of freedom is
intense. A number of nonlinear dynamical
effects have been found \cite{JETPL02,PRA07,PRA08} with point-like
atoms in a near-resonant rigid optical lattice:
chaotic Rabi oscillations, chaotic walking, dynamical fractals,
L\'evy flights, and anomalous diffusion.

Dynamical chaos in classical mechanics is a special kind of random-like motion
without any noise and/or random parameters. It is characterized by exponential
sensitivity of trajectories in the phase space to small variations in
initial conditions and/or control parameters. Such sensitivity does not
exist in isolated quantum systems because their evolution is unitary, and there is
no well-defined notion of a quantum trajectory. Thus, there is a fundamental
problem of emergence of classical dynamical chaos from more profound
quantum mechanics     which is known as quantum chaos problem   and the related
problem of quantum-classical correspondence.
In a more general context it is a problem of wave chaos. It is clear now that
quantum chaos, microwave, optical, and acoustic chaos
have much in common (see \cite{Makarov} for a review). The common practice
is to construct an analogue for a given wave object
in a semiclassical (ray) approximation and study its chaotic properties (if any) by
well-known methods of dynamical system theory. Then, it is necessary to solve
the corresponding linear wave equation in order to find manifestations of
classical chaos in the wave-field evolution in the same range of the control
parameters. If one succeeds in that a quantum-classical or wave-ray
correspondence is announced to be established.

In this paper we perform this program with cold two-level atoms in a one-dimensional
standing-wave laser field and show that coherent dynamics of the atomic matter waves
is really complicated in that range of the control parameters where the
corresponding classical point-like atomic motion can be strictly characterized as
a chaotic one. The effect is explained by a proliferation of atomic 
wave packets at the nodes of the standing wave.

\section{Order and chaos in dynamics of atomic wave packets in a laser
standing wave}

The Hamiltonian of a two-level atom, moving
in a one-dimensional classical standing-wave laser field,
can be written in the frame rotating with the laser frequency $\omega_f$ as follows:
\begin{equation}
\hat H=\frac{\hat P^2}{2m_a}+\frac{\hbar}{2}(\omega_a-\omega_f)\hat\sigma_z-
\hbar \Omega\left(\hat\sigma_-+\hat\sigma_+\right)\cos{k_f \hat X},
\label{Ham}
\end{equation}
where $\hat\sigma_{\pm, z}$ are the Pauli operators for the internal atomic degrees
of freedom, $\hat X$ and $\hat P$ are the atomic position and momentum operators,
$\omega_a$ and $\Omega$ are the atomic transition and Rabi frequencies, respectively.
We will work in the momentum representation and expand
the state vector as  follows:
\begin{equation}
\ket{\Psi(t)}=\int [a(P,t)\ket{2}+b(P,t)\ket{1}]\ket{P}dP,
\label{wavef}
\end{equation}
where $a(P,t)$ and $b(P,t)$ are the probability amplitudes to find atom
at time $t$ with the momentum $P$ in the excited and ground states, respectively.
The Schr\"odinger equation for these amplitudes is 
\begin{equation}
\begin{aligned}
i \dot a(p) &= \frac12 (\omega_rp^2 - \Delta)a(p) - \frac12[b(p+1) + b(p-1)],
 \\
i \dot b(p) &= \frac12 (\omega_rp^2 + \Delta)b(p) - \frac12[a(p+1)+ a(p-1)],
\end{aligned}
\label{Schp}
\end{equation}
where dot denotes differentiation with respect to dimensionless
time $\tau \equiv \Omega t$, and the atomic momentum $p$ is measured in units
of the photon momentum $\hbar k_f$.
The normalized recoil frequency, $\omega_r \equiv\hbar k_f^2/m_a\Omega$,
and the atom-field detuning,
$\Delta \equiv (\omega_f-\omega_a)/\Omega$, are the control parameters.

\begin{figure}[htb]\center
\includegraphics[width=0.45\textwidth,clip]{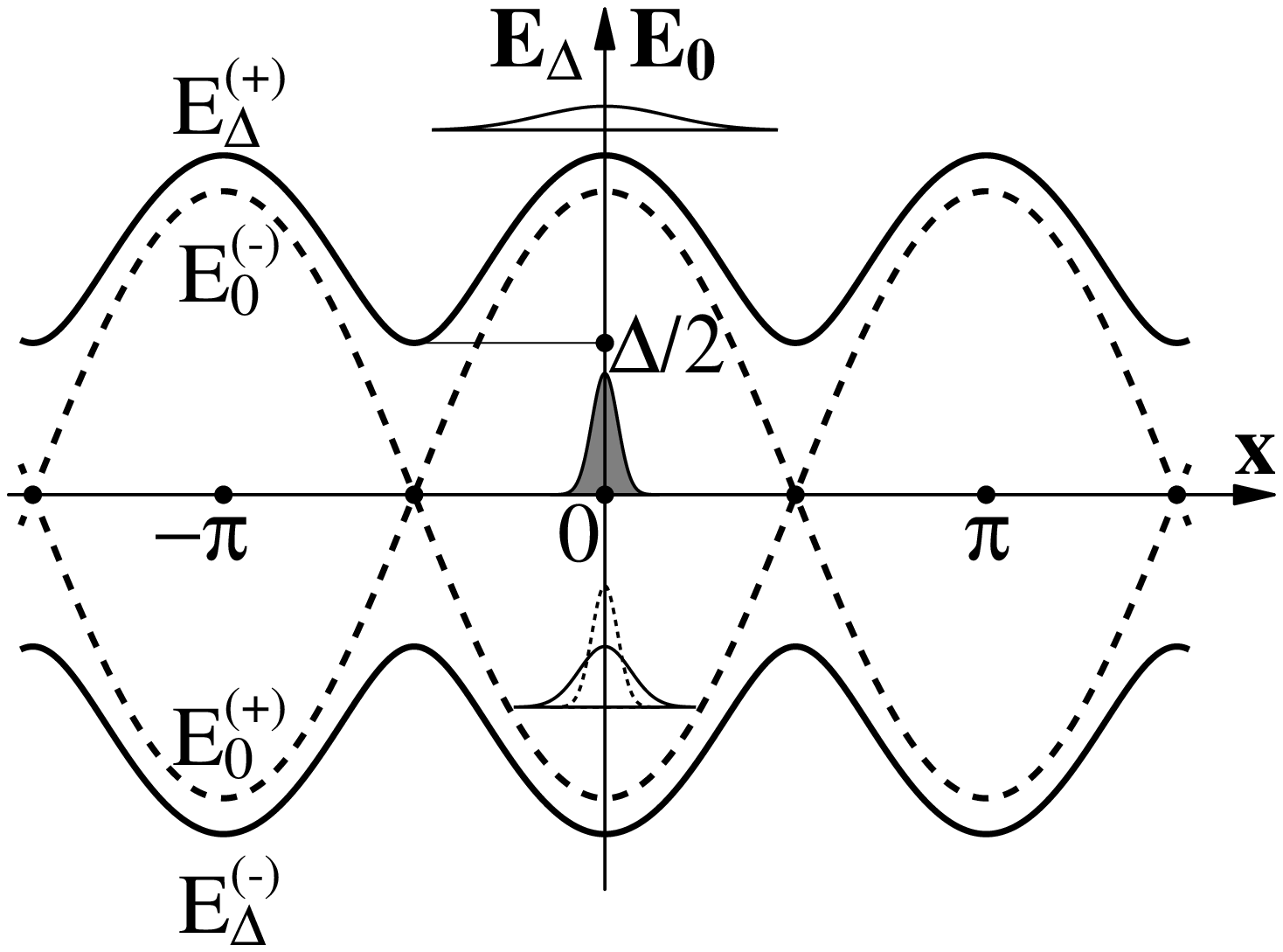}
\caption{The resonant
$E_0^{(\pm)}$ ($\Delta=0$, dotted curves) and non-resonant
$E_\Delta^{(\pm)}$ ($\Delta\not=0$, solid curves) potentials of a two-level atom
in a standing-wave laser field. An atomic wave packet, centered at
$x_0=0$, and its initial evolution in the upper and lower potentials are shown
schematically.}
\label{fig1}
\end{figure}

We will treat the wave-packet motion in the dressed-state
basis \cite{Cohen}
\begin{equation}
\ket{+}_\Delta = \sin{\Theta}\ket{2} + \cos{\Theta}\ket{1},\ 
\ket{-}_\Delta = \cos{\Theta}\ket{2} - \sin{\Theta}\ket{1}.
\label{dressf}
\end{equation}
The dressed states are eigenstates of an atom at rest in a laser field with
the eigenvalues of the quasienergy $E_\Delta^{(\pm)}$
and the mixing angle $\Theta$:
\begin{equation}
E_\Delta^{(\pm)} = \pm\sqrt{\frac{\Delta^2}{4} + \cos^2{x}},\ 
\tan{\Theta} \equiv \frac{\Delta}{2\cos{x}} - \sqrt{\left(\frac{\Delta}
{2\cos{x}}\right)^2 + 1}.
\label{dresse}
\end{equation}
The depth of the nonresonant optical potential is
\begin{equation}
U_\Delta=\left |\sqrt{\frac{\Delta^2}{4} +1} -\frac{|\Delta|}{2}\right |.
\label{depth}
\end{equation}
At $\Delta=0$, the depth changes abruptly from $1$ to $2$.
The resonant $E_0^{(\pm)}$ ($\Delta=0$) and non-resonant
$E_\Delta^{(\pm)}$ ($\Delta\not=0$) potentials of an atom
in a standing-wave field are drawn in Fig. \ref{fig1}.
The ground  atomic state can be  written as a superposition of
the dressed states
\begin{equation}
\ket{1}=\ket{+}_\Delta\frac{1}{\sqrt{1+\tan^2{\Theta}}} -
\ket{-}_\Delta\frac{\tan{\Theta}}{\sqrt{1+\tan^2{\Theta}}}.
\label{ground}
\end{equation}
In general, an atom in the ground state, placed initially at $x_0=0$,
will move along two trajectories simultaneously because it 
is situated simultaneously at the top of
$E_\Delta^{(+)}$ and  the bottom of  $E_\Delta^{(-)}$
(see Fig.\ref{fig1}).

In the dressed-state basis, the probability amplitudes to find the atom
at the point $x$ in the potentials $E_\Delta^{(+)}$ and  $E_\Delta^{(-)}$ are,
respectively,
\begin{equation}
C_+(x)=a(x)\sin{\Theta}+ b(x)\cos{\Theta},\
C_-(x)=a(x)\cos{\Theta}- b(x)\sin{\Theta},
\label{cprob}
\end{equation}
where the amplitudes in the bare-state basis, $a(x)$ and  $b(x)$, are
computed in the position representation with the help of the Fourier
transform
\begin{equation}
a(x)={\rm const}\int_{-\infty}^{\infty}dp'e^{ip'x}a(p'),\
b(x)={\rm const}\int_{-\infty}^{\infty}dp'e^{ip'x}b(p').
\label{Four}
\end{equation}

To clarify the character of the bipotential motion we write down 
the Hamiltonian of the internal degrees of freedom in the basis
$\ket{\pm}=(\ket{1} \pm \ket{2})/\sqrt{2}$: 
\begin{equation}
\hat H_{\rm int}=\hat \sigma_z \cos x + \frac{\Delta}{2} \hat \sigma_x.
\label{Hamdress}
\end{equation}
Let us linearize the potential in the vicinity of a node of the standing wave
and estimate a small distance the atom makes when crossing the node as follows:
$\delta x \simeq \omega_r |p_{\rm node}|\tau $, where
$|p_{\rm node}|$  is the mean atomic momentum near the node. 
The quantity $\omega_r|p_{\rm node}|$ 
is a normalized Doppler shift for an atom moving with the momentum 
$|p_{\rm node}|$, i.e., $\omega_D \equiv \omega_r|p_{\rm node}| \equiv k_f
|v_{\rm node}|/\Omega$. We arrive now at the famous
Landau--Zener problem \cite{LZ} to find a probability of transition 
between the states (\ref{dressf}) when the energy difference
varies linearly in time. In other words it is a a probability of transition 
from one potential to another or from one trajectory of motion to another. 
The asymptotic solution is 
\begin{equation}
P_{\rm LZ} = \exp \left(-\frac{\pi\Delta^2}{\omega_D}\right).
\label{LZ}
\end{equation}
There are three possibilities.

1. $\Delta^2 \gg \omega_D$. The transition probability 
is exponentially small and an atom moves
adiabatically along, in general, two trajectories without any transitions 
between them.

2.  $\Delta^2 \ll \omega_D$. The probability of a Landau-Zener
transition is close to unity, and an atom changes the potential each time upon
crossing any node, i.e., it moves mainly in the resonant potentials  $E_0^{(\pm)}$.

3. $\Delta^2 \simeq \omega_D$. The probabilities
to change the potential or to remain in the same one upon crossing a node are of
the same order. In this regime one may expect strong 
complexification of the wave function.

At large and small detunings, the translational motion splits into two 
independent motions in the potentials $E_{\Delta}^{(\pm)}$, and the wave-packet 
motion is regular in the first two cases. 
In the third case, the motion is complex because of a 
proliferation of wave packets at the nodes of the standing wave.  
We call such a motion as "a chaotic" one by the reasons which will be clear in 
section 3. The switch between the regular and chaotic regimes of atomic 
motion can be easily performed by changing the detuning.

Our normalization enables to change dimensionless values of the recoil frequency $\omega_r$
and the detuning $\Delta$ varying the Rabi frequency $\Omega$.
Working with a cesium atom at the transition $6S_{1/2}$ -- $6P_{3/2}$
($m_a =133$ a.~u.,
$\lambda_a = 852.1$ nm and $\nu_{\rm rec}\simeq2$ KHz \cite{Mlynek}), we
have $\omega_r=10^{-3}$ at $\Omega =1$ MHz. 
Let the atom is initially prepared in the ground state
as a Gaussian wave packet in the momentum space with $p_0=55$. 
The probability to find the atom with the momentum $p$ at time $\tau$ is
${\it P}(p,\tau)=|a(p,\tau)|^2 + |b(p,\tau)|^2$.

\begin{figure}[htb]
\begin{center}
\includegraphics[width=0.45\textwidth,clip]{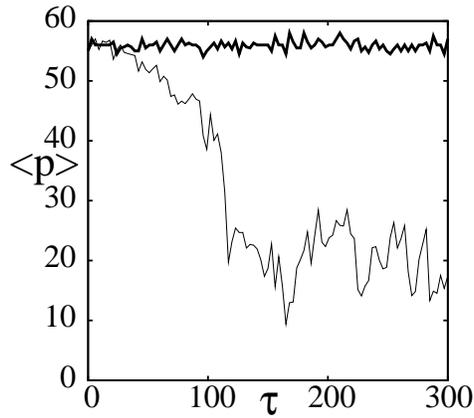}
\end{center}
\caption{Mean atomic momentum $\left<p\right>$ vs time
in the regular  ($\Delta=1$, the upper bold curve)
and chaotic  ($\Delta=0.2$) regimes of motion.}
\label{fig2}
\end{figure}

To illustrate the difference between the regular and chaotic regimes
of the wave-packet motion we take two values of the detuning, $\Delta =1$ and
$\Delta =0.2$. At $\Delta =1$, the motion is expected to be adiabatic and 
regular  because $\Delta^2 \gg \omega_D=0.055$, and the Landau--Zener
probability, $P_{\rm LZ}$, is exponentially small. At $\Delta =0.2$, 
one expects a much more complicated wave-packet motion with 
nonadiabatic transitions between the 
two potentials at the field nodes because $\Delta^2 =0.04 \simeq \omega_D$.
Figure~\ref{fig2} shows the dependence of the mean atomic momentum
$<p>$ over a large time scale in those two cases. In both the cases 
$<p>$ oscillates in a rather irregular way. The difference is that  
for an adiabatically moving wave packet, which we refer as a regular motion,  
the mean atomic momentum oscillates
in a narrow range around $p_0=55$ (the upper curve in the figure). 
Whereas the range of its 
oscillations for a wave packet moving with nonadiabatic transitions at 
the field nodes, which we refer as a chaotic motion, is much more broad 
(the lower curve in the figure). It is a simple illustration of the 
two different regimes of the wave-packet propagation in terms of 
the classical variable.

\begin{figure}[htb]
\begin{center}
\includegraphics[width=0.45\textwidth,clip]{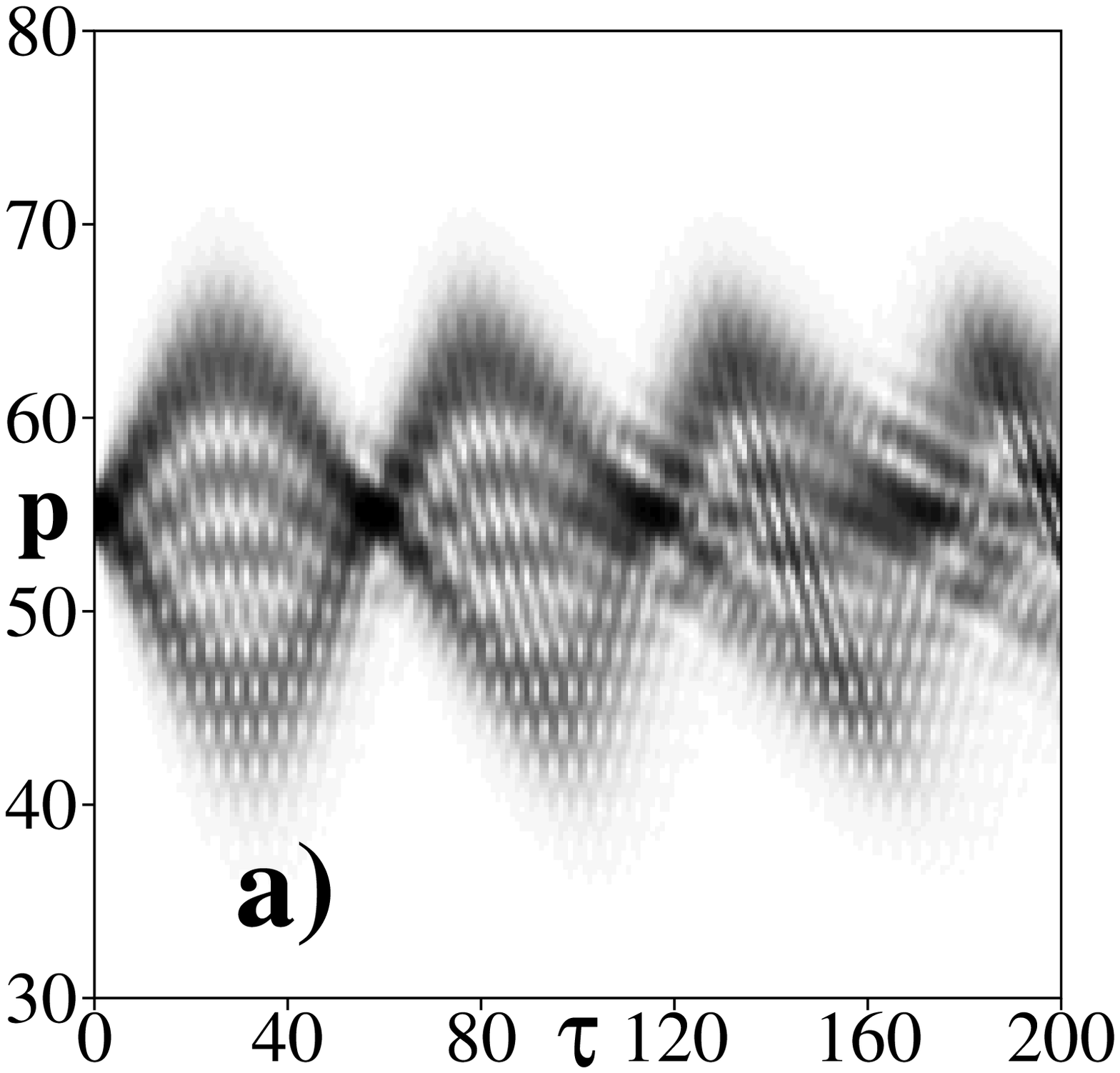}
\includegraphics[width=0.45\textwidth,clip]{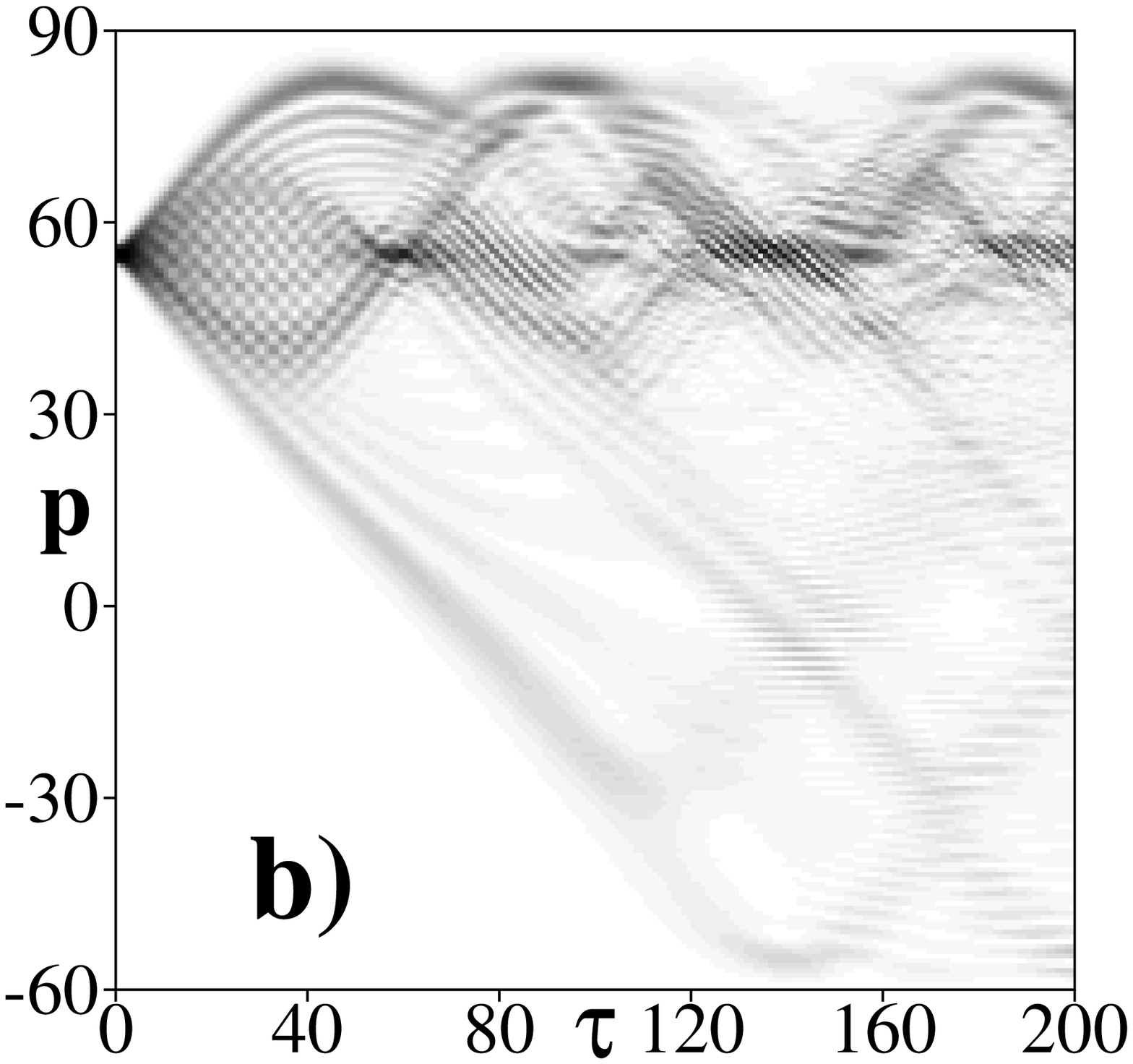}
\end{center}
\caption{Momentum probability-density  distribution
${\it  P}(p,\tau)$
vs time in (a) the regular  and (b) chaotic  regimes of motion.
The color codes the corresponding values of ${\it  P}(p)$.}
\label{fig3}
\end{figure}

In Fig.~ \ref{fig3}a we plot the dependence of the
momentum probability-density  on time at $\Delta =1$. The initial
wave packet splits from the beginning to a few components because the initial ground
state is a superposition of the dressed states (\ref{dressf}). The initial
kinetic energy is enough to perform a ballistic motion. The momentum
changes in a comparatively small range, from 40  to 70 of the
photon-momentum units. The packet does not split
at the nodes of the standing wave but it, on the contrary, recollects in
the momentum space at the nodes and spreads in between. However,
this recollection smears out in course of time.

At $\Delta=0.2$, the atomic ground state is the following superposition of
the dressed states: $\ket{1}\simeq 0.74\ket{+}_{\Delta}+ 0.66\ket{-}_{\Delta}$.
The $\ket{+}_{\Delta}$-component of the initial wave packet, i.e., that one,
starting from the top of the potential
$E_\Delta^{(+)}$, overcomes the barriers of that potential and  moves
in the positive  direction of the axis $x$ proliferating at the nodes.
As to the $\ket{-}$-component with decreased values of
$p$, it will be trapped in the potential well performing
oscillations in the momentum and position spaces. The period of those
oscillations is about  $T\simeq280$ which is equal approximately to the
period of revivals of the Rabi oscillations for the population inversion.
Figure \ref{fig3}b illustrates the effect of simultaneous trapping and
ballistic motion of the atomic wave packet in the chaotic  regime
resulting in a broad momentum distribution, from $p=-60$
to $p \simeq 80$.

\begin{figure}[htb]
\begin{center}
\includegraphics[width=0.45\textwidth,clip]{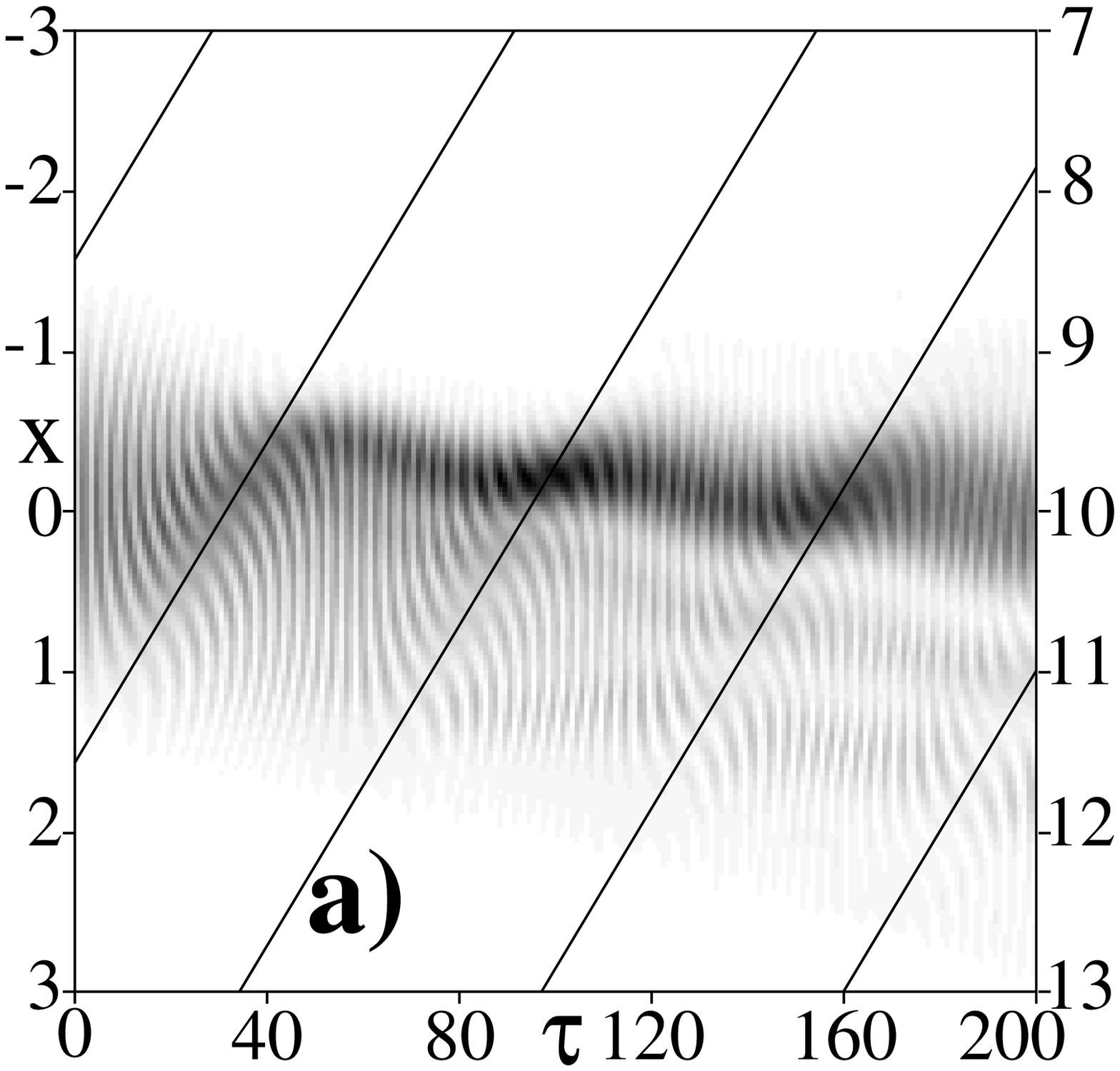}
\includegraphics[width=0.45\textwidth,clip]{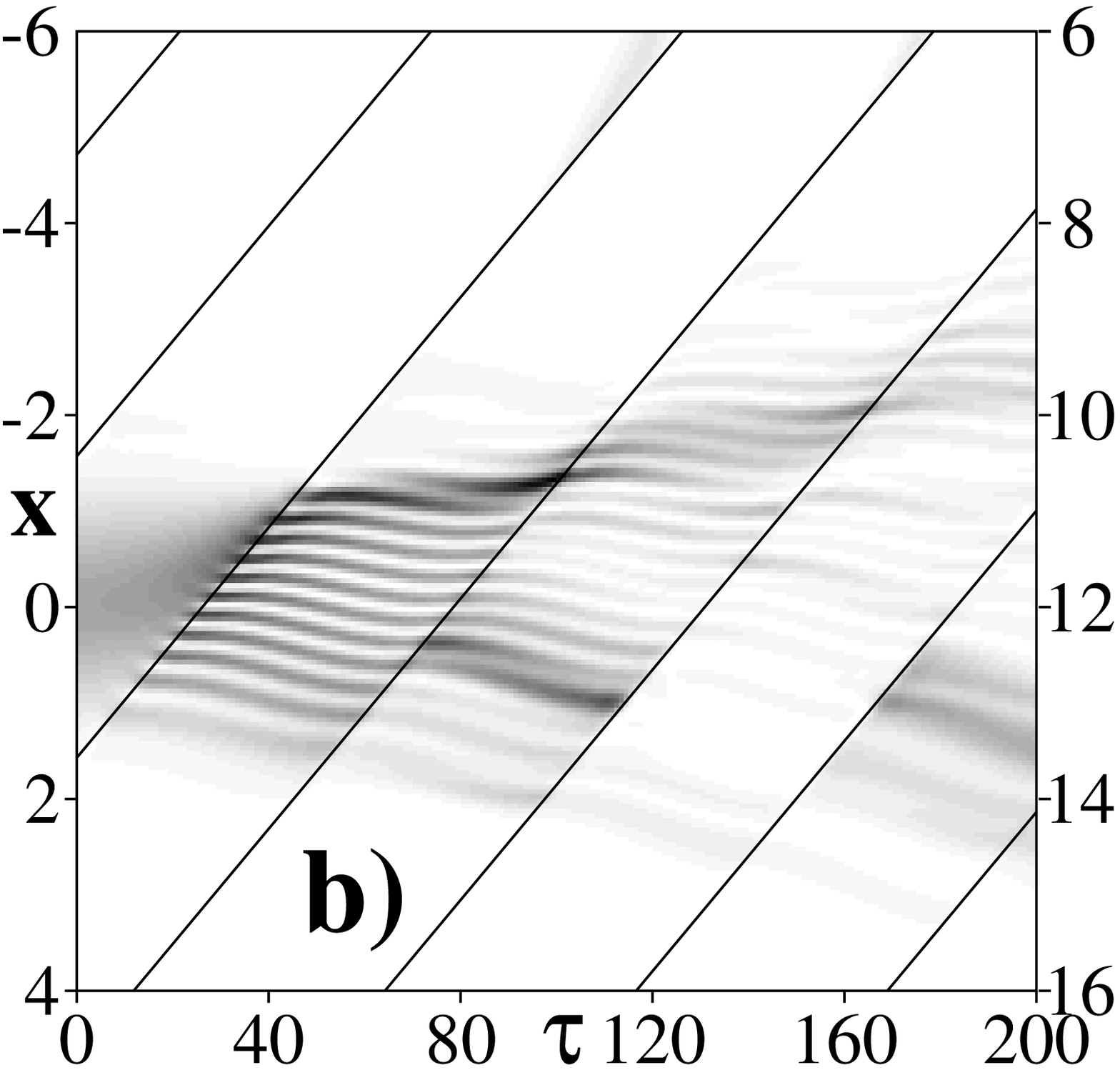}
\end{center}
\caption{The probability $|C_-(x)|^2$ to find the atom in the potential
$E_\Delta^{(-)}$ in the moving frame of reference in (a) the regular 
and (b)  chaotic regimes of motion. The slopes mark positions of the nodes
in the moving frame and the color codes the values of $|C_-(x)|^2$.}
\label{fig4}
\end{figure}

To illustrate the nonadiabatic transitions from one potential to another 
and their absence at the nodes more
explicitly, we go to the position space and compute the probabilities
$|C_\pm(x,\tau)|^2$ (\ref{cprob}) to be at the point  $x$ at time $\tau$
in the potentials $E_\Delta^{(+)}$ and $E_\Delta^{(-)}$, respectively.
In Fig.\,\ref{fig4} we plot the evolution of the probability density 
$|C_-(x,\tau)|^2$ in the frame moving with the initial atomic velocity
$\omega_rp_0=0.055$. The slope straight lines in the figure mark positions of
the nodes in the moving frame. In the case of the
regular  motion at $\Delta=1$ (Fig.~\ref{fig4}a), no transitions happen
when the atom crosses the nodes. In the chaotic  regime at  $\Delta=0.2$,
one observes visible changes in the
probability-density $|C_-(x)|^2$ exactly at the node lines (see 
Fig.~\ref{fig4}b). It means transitions from
one trajectory to another at the field nodes that should occur
in a specific range of the control parameters if
$\Delta^2 \simeq \omega_D$. This results in a proliferation  of
the components of the wave packet at the nodes and, therefore,
a complexification of the wave function (see Fig.~\ref{fig3}b).

\begin{figure}[htb]
\begin{center}
\includegraphics[width=0.45\textwidth,clip]{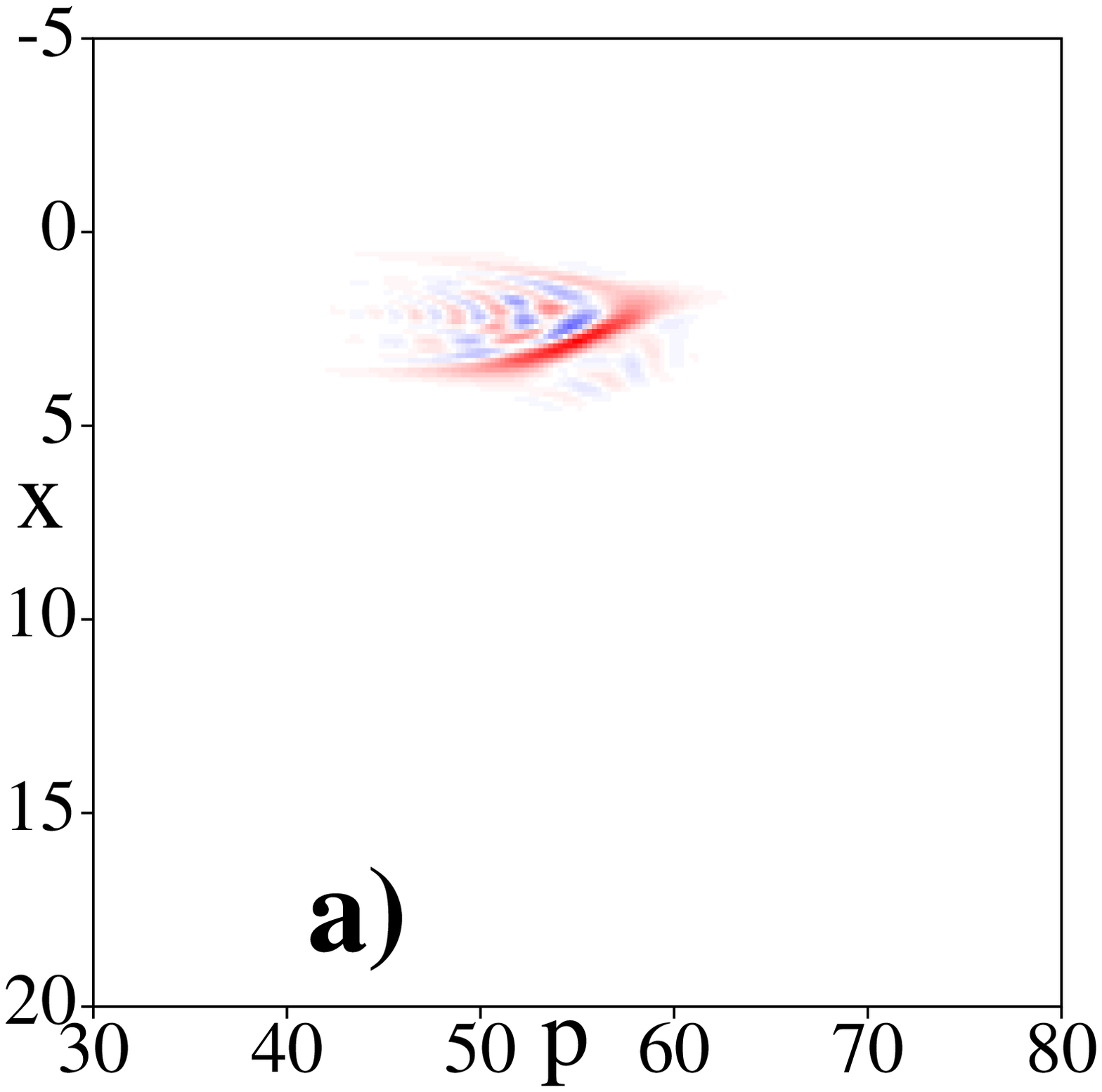}
\includegraphics[width=0.45\textwidth,clip]{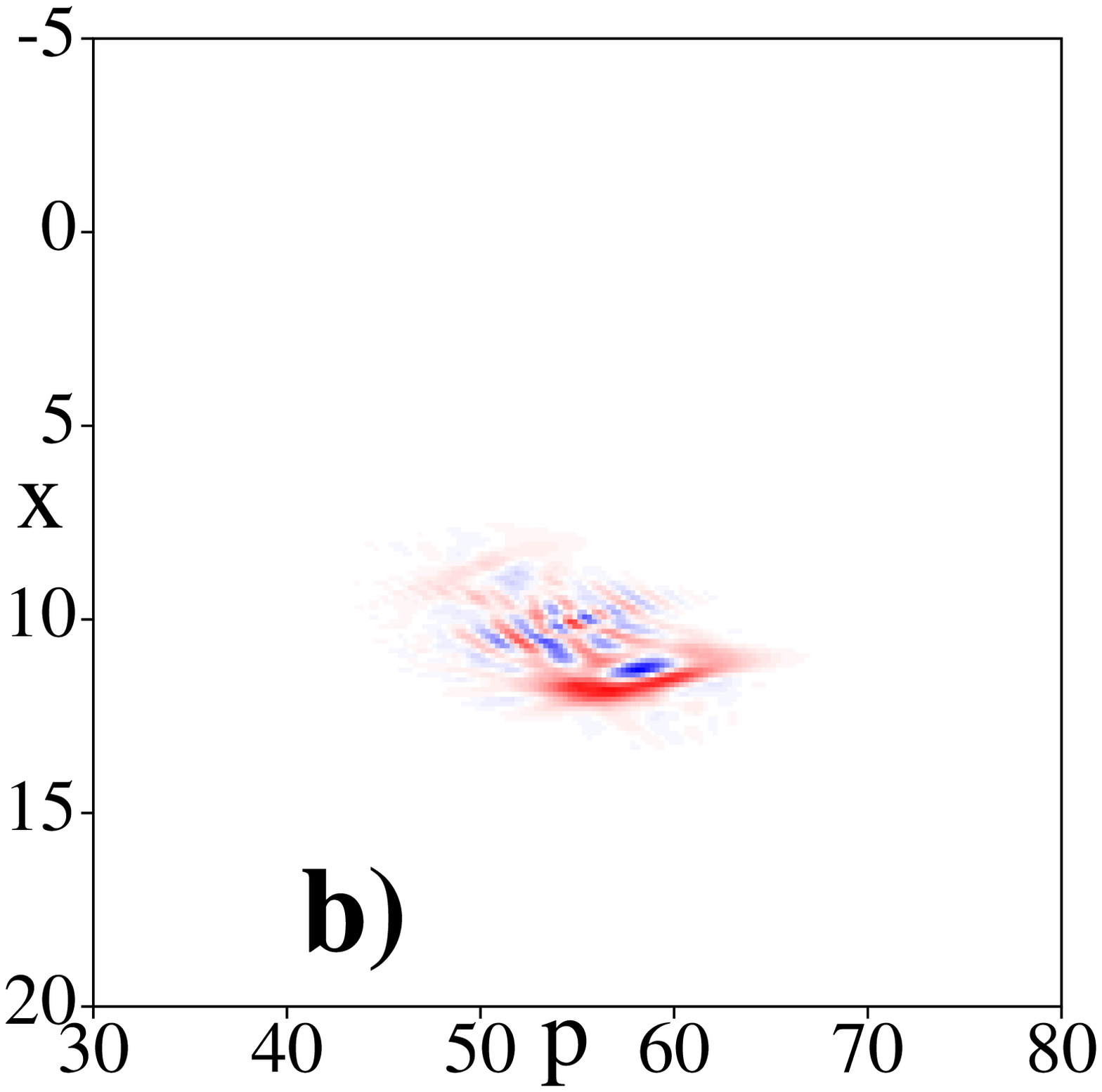}
\end{center}
\caption{Contour plots of the Wigner function at (a) $\tau=50$ 
and (b) $\tau=200$
for the regular  regime of the atomic motion. Color online: 
red and blue areas show 
positive and negative values of the Wigner function, respectively.}
\label{fig5}
\end{figure}

\begin{figure}[htb]
\begin{center}
\includegraphics[width=0.45\textwidth,clip]{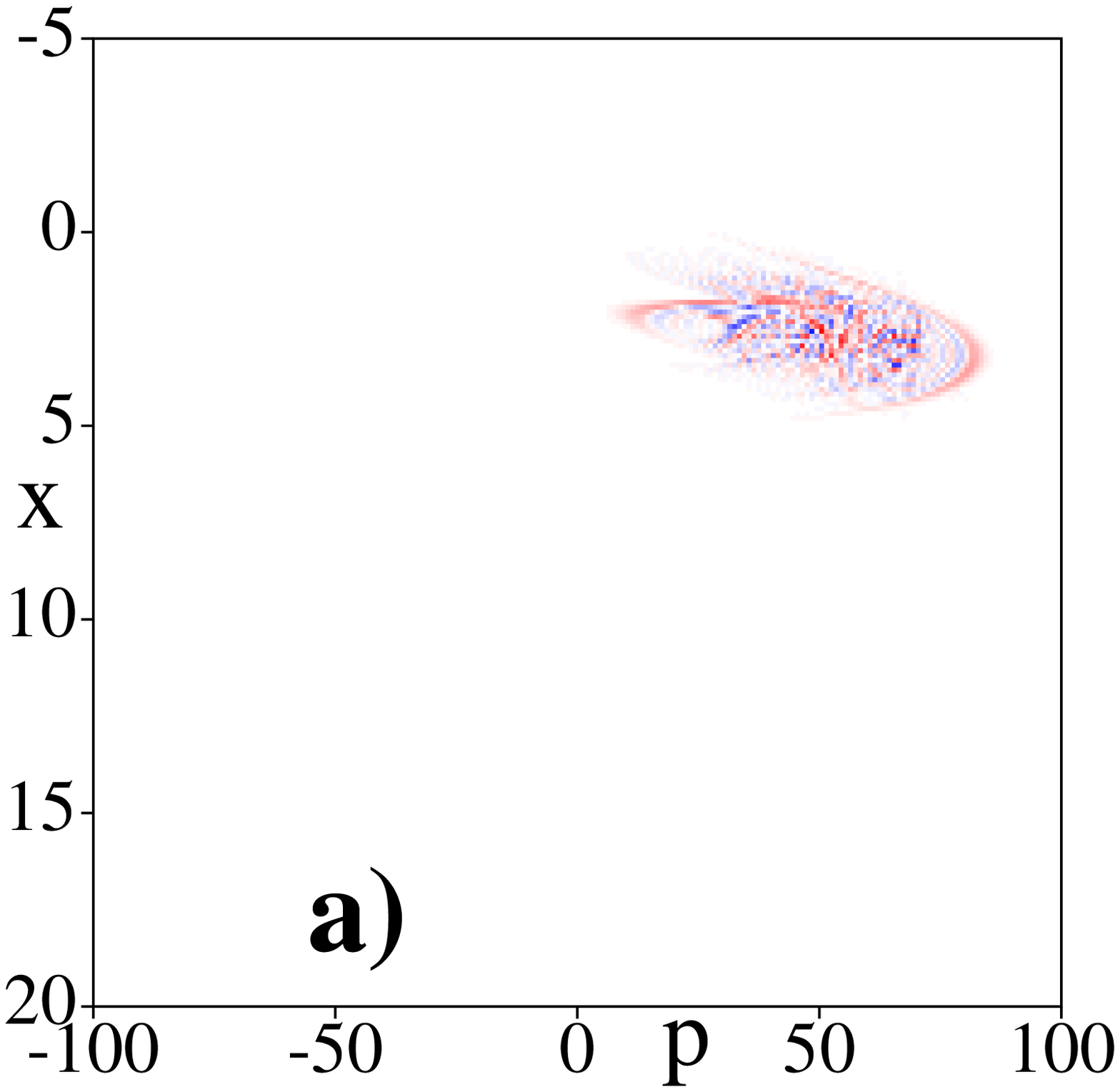}
\includegraphics[width=0.45\textwidth,clip]{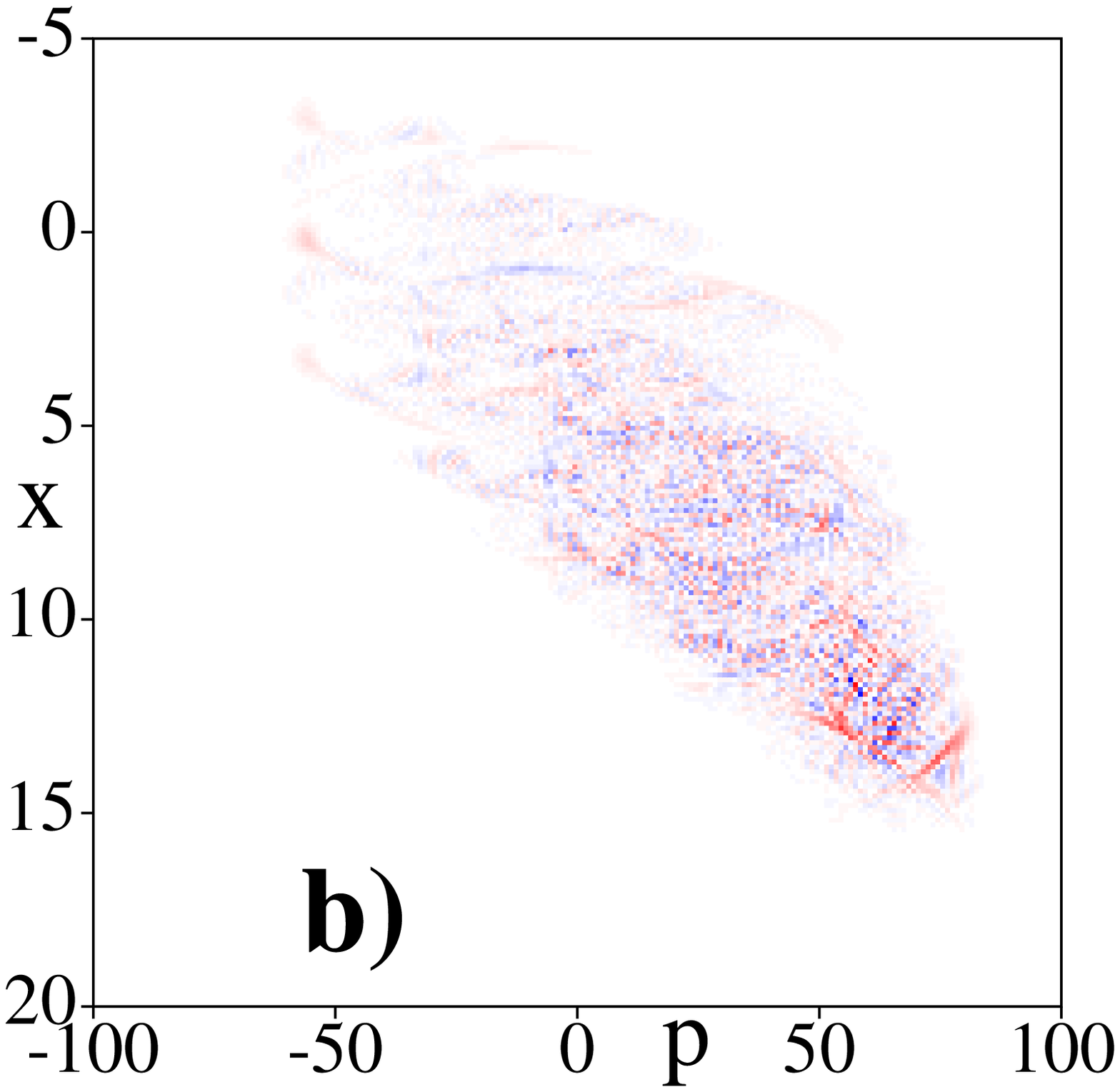}
\end{center}
\caption{The same as in Fig. \ref{fig5} for
the chaotic  regime of motion.}
\label{fig6}
\end{figure}

The Wigner function can be used to visualize
complexity of the wave function in the chaotic  regime of the atomic motion.
We compute the evolution of the Wigner function of the ground state in
the momentum space
\begin{equation}
W_b(p,x, \tau)={\rm const}\int_{-\infty}^{\infty}dp'e^{-ip'x}a(p-p'/2)
a^*(p+p'/2).
\label{Wigner}
\end{equation}
This quantity gives a quasi-probability distribution corresponding to a
general quantum state (\ref{wavef}). Figure~\ref{fig5} shows a contour plot of the Wigner
function (\ref{Wigner}) at two moments of time,  $\tau=50$ and $\tau=200$,
when the atom moves in a regular way ($\Delta=1$).   Figure~\ref{fig6} is
a contour plot of this function at the same times,                
but with an atom making nonadiabatic transitions  
at the field nodes ($\Delta=0.2$).
In the chaotic  regime (Fig.~\ref{fig6}b)
we see a dust-like distribution of nonzero values of the Wigner function
at $\tau=200$ which occupy much more larger area in the phase space than
the function for the regular  motion (Fig.~\ref{fig5}b).

\section{Quantum-classical correspondence}

In this section we compare the results of the quantum treatment with those
obtained for the same problem in the semiclassical approximation when the
translational motion has been treated as a classical one
\cite{JETPL02,PRA07}. We must compare quantum
results for a single atomic wave packet not with a single point-like atom but
with an ensemble of point-like  atoms. Dynamical chaos has been found and
analyzed in detail in Refs. \cite{JETPL02,PRA07} in the semiclassical
approximation. Both
the internal and external degrees of freedom of a two-level atom in a standing
wave field have been shown to be chaotic in a specific range of values of
the  detuning $\Delta$, the recoil frequency $\omega_r$, and the initial 
momentum $p_0$. 

Coherent semiclassical evolution of a point-like 
two-level atom is governed by the Hamilton-Schr\"odinger equations with the same 
normalization as in the quantum case \cite{PRA07}
\begin{equation}
\begin{gathered}
\dot x=\omega_r p,\quad \dot p=- u\sin x, \quad \dot u=\Delta v,
\\
\dot v=-\Delta u+2 z\cos x, \quad
\dot z=-2 v\cos x,
\end{gathered}
\label{mainsys}
\end{equation}
where 
\begin{equation} 
u\equiv2 \Rre(a_0b_0^*),\ v\equiv-2\Iim(a_0b_0^*),\ z\equiv|a_0|^2-|b_0|^2
\label{uvz}
\end{equation}
are the atomic-dipole components ($u$ and $v$) and population-inversion ($z$), 
and $a_0$ and $b_0$ are the complex-valued probability amplitudes to find the
atom in the excited, $\ket{2}$, and ground, $\ket{1}$, states, respectively. 
The system (\ref{mainsys}) has two integrals of motion, the total energy
\begin{equation}
W\equiv\frac{\omega_r}{2}p^2-u\cos x-\frac{\Delta}{2}z,
\label{H}
\end{equation}
and the length of the Bloch vector, $u^2+v^2+z^2=1$. 
Equations (\ref{mainsys}) constitute a nonlinear Hamiltonian
autonomous system with two and half degrees of freedom and two 
integrals of motion. 

\begin{figure}[htb]\center
\includegraphics[width=0.45\textwidth,clip]{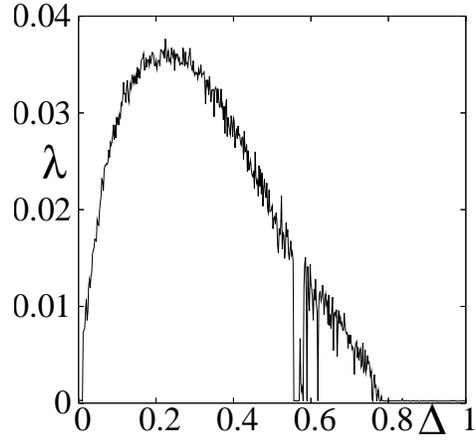}
\caption{Maximal Lyapunov exponent $\lambda$, computed with semiclassical equations
of motion (\ref{mainsys}), vs the detuning $\Delta$ at $\omega_r=10^{-3}$ and 
$p_0=55$.}
\label{fig7}
\end{figure}

In classical mechanics there is a qualitative  criterion of dynamical
chaos, the maximal Lyapunov exponent $\lambda$, which measures the mean rate
of the exponential divergence of initially closed trajectories in the phase
space.  In Fig.~\ref{fig7} we plot this quantity, computed with semiclassical 
equations of motion (\ref{mainsys})  vs
the detuning $\Delta$ at $\omega_r=10^{-3}$ and $p_0=55$. 
At zero detuning, the set of 
semiclassical equations acquires an additional integral of motion and becomes 
integrable. The center-of-mass motion is regular at small values of the 
detuning, $\Delta \ll 1$, and at large ones, $\Delta >0.8$. 
Positive values of $\lambda$ at $0<\Delta <0.8$ characterize  unstable  motion.
Local instability of the center-of-mass motion produce chaotic
motion of an atom in a rigid standing wave without any modulation of its
parameters in difference from the situation with atoms in a periodically
kicked optical lattice \cite{Raizen,HH01}. There is a range of
initial conditions and the control parameters where the center-of-mass
motion in an absolutely deterministic standing wave resembles a random
walking. It means that a point-like atom alternates between
flying through the lattice, and being trapped in its wells changing
the direction of motion in a random-like way \cite{PRA07,PRA08}.

\begin{figure}[htb]\center
\includegraphics[width=0.85\textwidth,clip]{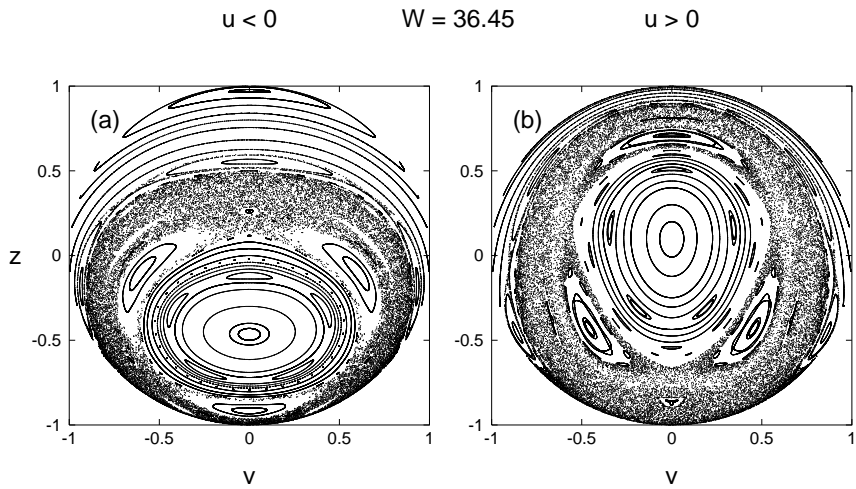}
\caption{Poincar\'e sections of the Bloch
sphere illustrating the effect of semiclassical chaos 
with point-like atoms at $\omega_r=10^{-5}$, $\Delta = -0.05$ and 
the total atomic energy $W=36.45$. 
(a) $u<0$ (western Bloch hemisphere), (b) $u>0$
(eastern Bloch hemisphere).}
\label{fig8}
\end{figure}

In Fig.~\ref{fig8} we illustrate semiclassical chaos with 
the Poincar\'e mapping for a number of atomic ballistic trajectories 
in the western ($u<0$) and eastern ($u>0$) hemispheres of the Bloch 
sphere $(u,v,z)$ on the plane $v-z$. One can see 
a typical structure with regions of regular motion in the 
form of islands and chains of islands filled by regular trajectories. 
The islands are imbedded into a stochastic sea, and they are produced by nonlinear resonances of
different orders. Increasing the resolution of the mapping, one can see
that large islands are surrounded by islands of a smaller size each of
which, in turn, is surrounded by a chain of even more smaller islands, and so on.

In quantum mechanics there is no well-defined notion of a trajectory in the
phase space, the very phase space is not continuous due to the Heisenberg
uncertainty relation, and, hence, the Lyapunov exponents can not be
computed (however, see Ref. \cite{Manko} where a notion  of a Lyapunov exponent 
for quantum dynamics has been discussed). The main result of this paper is the establishment of the fact,
that chaotically-like complexification of the wave function, 
caused by nonadiabatic transitions at the field nodes, occurs exactly 
at the same range of the control parameters where the semiclassical
dynamics has been shown to be chaotic in Refs. \cite{JETPL02,PRA07}.
It should be stressed that quantum motion of a wave
packet with nonadiabatic transitions between  the two optical potentials is
compared with the center-of-mass motion of an ensemble of atoms each of which
moves in a single optical potentials. So, when we say about a
quantum-classical correspondence we mean a correspondence between the wave
function of a single quantum atom and the trajectories of the ensemble of
classical atoms with different values of the initial momentum $p_0$ and at the
other equal conditions.

\section{Conclusion}

We have studied coherent dynamics of cold atomic wave  packets in
a one-dimensional standing-wave laser field. The problem has been considered
in the momentum representation and in the dressed-state basis where the motion
of a two-level atom was interpreted as a propagation  in two optical potentials.
The character of that motion has been shown to depend strongly on the ratio of
the squared detuning, $\Delta^2$, to the  normalized Doppler shift
$\omega_D$. In the regular  regime, when $\Delta$ is comparatively large or small,
wave packets move in a simple way. The chaotic  regime occurs
if $\Delta^2\simeq \omega_D$ when the probability for an atom to make nonadiabatic
transitions while crossing the nodes of the standing wave is large. Atom in this regime
of motion simultaneously moves ballistically and is trapped in a well of the optical
potential. This type of motion and
proliferation of wave packets at the nodes result in a complexification
of the wave function both in the momentum and position spaces manifesting
itself in the irregular behavior of the Wigner function. 

Comparing the
results of the quantum treatment with those obtained
in the semiclassical approximation, when the
translational motion has been treated as a classical one
\cite{JETPL02,PRA07}, we have found that the wave-packet dynamics is
complicated exactly in that range of
the  atom-field detuning and recoil frequency where the classical
center-of-mass motion has been shown to be chaotic in the sense of exponential
sensitivity to small variations in initial conditions or parameters.

As to possible practical applications of the results obtained we mention   
atomic lithography to produce small-scale complex prints of cold atoms 
(see, for example, beautiful experiments on coherent matter-wave manipulation 
\cite{ZC04,ZL09,WT09}), new ways to manipulate atomic motion in optical 
lattices by varying the atom-filed detuning and atomic ratchets with cold atoms.

\section*{Acknowledgments}
This work was supported  by the Russian Foundation for Basic Research
(project no. 09-02-00358), the Integration grant from the Far-Eastern
and Siberian branches of the RAS, and  the Program
``Fundamental Problems of  Nonlinear Dynamics'' of the RAS. I would like 
to thank L. Konkov for preparing Figs.6 and 7.


\end{document}